\documentclass[aps,pra,preprint,floatfix]{revtex4-1}
\usepackage{graphicx,xcolor}
\usepackage{amsmath,amssymb}
\usepackage{dsfont}

\newcommand{\heff}{H_{\scriptscriptstyle \rm eff}}
\newcommand{\hs}{\mathcal{H}_{\scriptscriptstyle \rm S}}
\newcommand{\hcs}{\mathcal{H}_{\scriptscriptstyle \rm CS}}
\newcommand{\hdr}{\mathcal{H}_{\scriptscriptstyle \rm DR}}

\newcommand{\hl}{\mathcal{H}_{\scriptscriptstyle \rm L}}

\newcommand{\hsl}{\mathcal{H}_{\scriptscriptstyle \rm SL}}
\newcommand{\mth}{M_{\scriptscriptstyle \rm th}}
\newcommand{\wsl}{\omega_{\scriptscriptstyle \rm SL}}

\newcommand{\hdd}{\mathcal{H}_{\scriptscriptstyle \rm DD}}

\newcommand{\nn}{\nonumber}

\newcommand{\tr}{\text{Tr}}
\newcommand{\trl}{\text{Tr}_{ \scriptscriptstyle \rm L}}
\newcommand{\rh}{\rho_{\scriptscriptstyle  S}}
\newcommand{\rl}{\rho_{\scriptscriptstyle \rm L}^{\scriptscriptstyle \rm eq}}

\newcommand{\etal}{\textit{et al.\ }}

\begin{document}

\title{ Determination of Fluctuation correlation time in solid-state Nuclear Magnetic Resonance }

\author{Saptarshi Saha}
\email{s.saha@tu-berlin.de}
\affiliation{Institut f{\"u}r Physik und Astronomie, Technische Universität Berlin, Hardenbergstr. 36, 10623 Berlin, Germany}
\author{Rangeet Bhattacharyya}
\email{rangeet@iiserkol.ac.in}
\affiliation{Department of Physical Sciences, Indian Institute of Science
Education and Research Kolkata, Mohanpur - 741 246, West Bengal, India}
 
\begin{abstract}

Solid-state NMR provides a wide variety of experimental techniques to detect and analyze
a material's chemical and physical environment. Here, we offer a theoretical demonstration of a
new approach that could be a promising candidate for characterizing the local
environments and shifts. Prethermalization by applying a spin-locking pulse
brings a new research paradigm in solid-state NMR and has not been hitherto explored
for such purposes. We show that a prethermal state can also be effective in this case. A
prethermal state is described using its lifetime and the value of transverse
magnetization. Using these two variables, we successfully detect the changes in the
environmental parameter and chemical and dynamical shifts (such as Lamb shifts). Our results exhibit that the lifetime
increases with increasing environmental correlation time. On the other hand, the
transverse magnetization decreases with the increase in the strength of the shift parameter. Based
on these observations, we propose that the prethermalization dynamics can yield important
information on local environment.

\end{abstract}
 
\maketitle 

\section{Introduction}

Since the discovery of nuclear magnetic resonance (NMR) spectroscopy \cite{Bloch1946}, the
experimental technique has had enormous applications in several branches of science
\cite{Duer2002, abragam1961principles}. The ability of NMR to probe the atomic nuclei
using an external field revolutionized the way to demonstrate the structure of a material
at the microscopic level \cite{muller_twodimensional_1975, Luciano1974, Nagayama1980}.  In
medical science, similar techniques are applied (e.g., magnetic resonance imaging (MRI))
as a diagnostic tool for high-resolution imaging of soft tissues \cite{Lee2005MRS}. NMR is
also used for studying the properties of pharmaceutical ingredients \cite{Pellecchia2002,
Pellecchia2008}. Similarly, in biochemistry, organic chemistry, and material science,
experimental schemes related to solid-state and liquid-state NMR play a pivotal role in
exploring molecular structure, chemical characterization, probing reaction mechanisms, and
intermolecular interactions \cite{Xu2019, Chakraborty2018}.  On the same note, NMR
provides a unique platform for quantum information processing and quantum technology
\cite{Jones2011}. Several pulse protocols in NMR have been utilized for precise control of
quantum states, generation of quantum entanglement, and quantum error correction up to a
significant accuracy \cite{Vandersypen1037}. Below, we mention one of the primary uses of NMR in
the quantum world.

 The nitrogen-vacancy (NV) center is known as a point defect in a diamond lattice composed
of $^{13}$C nuclei. Using NMR, one can excite the whole system by an external drive to
study various interesting properties \cite{Ajoy2015, Pagliero018, Ajoy2019}. Such a
platform recently gained a lot of attention in quantum computation and is also a topic of
interest in this paper.  Notably, the NV center is used to hyperpolarize the $^{13}$C
nuclei in a diamond lattice. The dynamic nuclear polarization (DNP) technique is often
used in NMR for such scenarios \cite{Khandelwal2023, Poggiali2017}. Those hyperpolarized
spins are very efficient for studying several non-equilibrium phenomena in many-body systems, quantum physics \cite{Beatrez2023b, Sahin2022}. For example, if the spins are initially
oriented along the `X' direction, a spin-locking pulse is applied to create a long-lived
prethermal state. The spin-locking pulse is defined as an on-resonance periodic excitation
applied along the transverse direction of the system \cite{Chakrabarti_2023}. By applying
the pulse, the spins can be locked in that particular direction in the intermediate
timescale (i.e, prethermalization) \cite{saha2023}. Such a prethermal state is very
significant in quantum computation as it has a long transverse relaxation time; hence, it
can hold the quantum memory (i.e., quantum coherence) for a longer time. The prethermal
state is also used for quantum sensing, the creation of discrete-time-crystalline phase,
quantum technologies, etc \cite{Beatrez2023, Saha2024,saha2024dissipative}. However, since
the first work by Redfield in that direction \cite{Redfield1955}, there was no specific
theory to explain such phenomena in a simplified manner for a long time.  

Recently, Chakrabarti \etal proposed an open quantum system formalism (i.e.,
fluctuation-regulated quantum master equation (FRQME) \cite{chakrabarti2018b}) to derive a
dynamical equation to describe such prethermalization phenomena \cite{Chakrabarti_2023}.
Those theoretical results show an excellent agreement with the experimental outcomes
\cite{beatrez_floquet_2021}. Saha \etal further expanded the theoretical prescription to
explain the decay of the prethermal state, which is also known as the $T_{1 \rho}$ process in
NMR \cite{saha2023}. 

In this manuscript, we mainly focus on characterizing the local environment using NMR.
We note that among the various existing tools, NMR
has been proven to be a powerful technique for analyzing atomic and molecular structures.  NMR is
used to probe the magnetic environment of atomic nuclei—most commonly hydrogen ($^1$H) and
carbon ($^{13}$C). Such probing is essential for gaining insights into molecular
connectivity, functional groups, stereochemistry, organic chemistry, pharmaceutical
research, biopolymers, and materials science \cite{DAgostino2017}.

Naturally, a question comes next: can such a spin-locking technique be used for the
characterization of the local environment?  Recently, by measuring $T_{1 \rho}$ using a spin-locking pulse
sequence, chemical characterization of solid pharmaceuticals is reported
\cite{Almeida2024}. However, none of the works have implemented the characterization
scheme using a long-lived prethermal state. In this manuscript, we predict that a
prethermal state will also be beneficial for such characterization. We take the
analytical route to show the performance of the prethermal state to differentiate the
chemical properties of solid-state NMR. We mainly focus on two aspects: the property of
the chemical environment and the presence of chemical shift.

Chemical environment and chemical shift are two critical factors that can directly
influence the NMR response. The chemical environment refers to the electronic
surroundings, which directly affect the line shape, Lamb shift, and relaxation rates
\cite{ernst2023,bloembergen_relaxation_1948,bloembergen_nuclear_1947}. On the other hand,
the chemical shift depicts the deviation from any ideal crystalline structure (e.g.,
vacancies, local disorder) due to the presence of external spins. Those shifts are helpful in analyzing the chemical properties of the material, conductivity, and changes in
polarization, etc. Hence, the characterization of the above-mentioned elements is
necessary in NMR for various purposes \cite{Jameson1996}.

In this manuscript, we provide an analytical demonstration of how to characterize the
chemical components using the recently explored prethermal state in solid-state NMR. There
exists a series of publications from the same author on the prethermalization using
spin-locking \cite{Chakrabarti_2023, saha2023, Saha2024}. Here, we adopt a similar
theoretical protocol used in those works to exhibit our main result in this manuscript. We
find that such characterization is possible using the prethermal state by measuring its
lifetime and the quasi-steady state value of transverse magnetization.

We arrange the manuscript in the following order.  In Sec. \ ref {sec-II}, we describe the
system, and the relevant Hamiltonian of the system is given. The corresponding dynamical
equation is given in Sec. \ ref {sec-III}, where we also discuss the presence of different
relaxation times in the system. A numerical solution of the above-mentioned differential
equation is also given in the same section. Using the solution, in Sec- \ref{sec-IV}, we
show how such a spin-locking pulse can be used for characterization. We mainly focus on
the two scenarios of characterization, e.g., the chemical environment and the chemical
shifts. Finally, we discuss our results, draw further implications in Sec-\ref{sec-V}, and
conclude in \ref{sec-VI}.

\section{Description of the system}
\label{sec-II}
 To mimic the system consisting of $^{13}$C atoms in the presence of an on-resonance periodic drive, we consider a rudimentary model of dipolar-coupled spin-half particles externally coupled to the environment. For simplicity, we only consider a two-spin ensemble in our analysis; hence, it represents a dilute sample. A spin-locking pulse is also applied in the `$X$' direction. For modeling the effect of the local electronic environment, we assume a Jaynes-Cummins type Hamiltonian, where the spins are coupled with environmental degrees of freedom \cite{breuer2002}. Furthermore,  we add a chemical shift Hamiltonian in our model \cite{smithii}. The total Hamiltonian of the system + environment is given by,
\begin{eqnarray}\label{ham}
\mathcal{H}(t)&=& \hs^{\circ} + \hl^{\circ} +  \hdr(t)+ \hcs + \hdd  + \hsl + \hl(t).
\end{eqnarray}
We give the details on the individual terms of the Hamiltonian below.
\begin{enumerate} 
\item  $\hs^{\circ}$
represents the Zeeman interaction and the analytical form is given by $\sum\limits_{i=1}^2\omega_\circ I_z^i$. Here, $\omega_\circ$ is the Zeeman frequency, and $I_{\alpha}^i = \sigma_{\alpha}^i/2, \,\alpha \in \{x,y,z\}$,
$\vec{\sigma}$ is the Pauli spin matrix. $i$ is the
spin index. 
\item The free Hamiltonian of the environment is $\hl^{\circ} = \omega_L L_+
L_-$, where $\omega_L$ is the frequency and $L_{\pm} $ are the corresponding creation and
annihilation operators of the local environment of each spin. The environment is always in equilibrium with a constant temperature (the inverse temperature is defined as $\beta$). Hence, the density matrix of the environment is given by, $\rl = e^{- \beta \hl^{\circ} }/\mathcal{Z_L}$. Here, $Z_L$ is the partition function. The environment is also assumed to be isotropic in nature. Hence, $\tr_L (\rl L_\pm ) = 0$.
\item $\hdr(t)$ represents the spin-locking pulse (i.e., an on-resonant periodic excitation applied to the spins along `$X$' direction \cite{saha2023},  $\hdr(t) = \sum\limits_{i=1}^2 \omega_1 I_{x}^i  \cos\omega
t$, with $\omega = \omega_\circ$.
\item The motion of the electron cloud around the nuclear spins results in a localized field, which causes a shift to the energy levels of the non-interacting spins  \cite{smithii}. The corresponding Hamiltonian is denoted by $\hcs$.  For simplicity, we only consider the `Z' component of the Hamiltonian, which is written as  $\hcs = \alpha I_z^i$. $\alpha$ denotes the amplitude of the shift.
\item The dipolar coupling between spins in terms of spherical tensors is given  $\hdd = \sum\limits_{m=-2}^{2}
(-1)^m \omega_{d_{m}} \mathcal{T}^{m}_2$ \cite{Duer2002}, where, $\omega_{d_{m}}$ is the coupling amplitude $\propto \mathcal{Y}^{-m}_2(\theta,\phi)/r^3 $. $\mathcal{Y}^{-m}_2(\theta,\phi)$ is the spherical harmonics
of rank $2$, and $\mathcal{T}^{m}_2$ is the irreducible spherical tensor (of rank $2$ and order $m$).
$(\theta,\phi)$ are the polar and azimuthal angles of the average orientation of the dipolar vector, and $\vec{r}$ is the average distance between the two spins.
\item  Following the Jaynes-Cummings model,
the system-bath coupling  $\hsl$ can be written as, $
\hsl = \sum\limits_{i=1}^2 \wsl (I_+^i L_-^i + h.c.)$.
Here, $\omega_{\text{\tiny{SL}}}$ is the system-local environment coupling amplitude \cite{breuer2002}. 
\item $\hl(t)$
represents the thermal fluctuation in the local environment \cite{chakrabarti2018b}. The analytical form is chosen as, $\hl(t) = \sum_i f_i(t)\vert \phi_i
\rangle \langle \phi_i \vert$, where, $\{\vert\phi_i\rangle\}$ are the eigen basis of the environment.  $f_i$
are the independent Gaussian stochastic variables with zero mean and standard deviation $\kappa \,
\left(\kappa^2=\frac{1}{\tau_c}\right)$, and $\tau_c$ is the correlation time. The chemical environment can be characterized by measuring $\tau_c$. We note that such a form of 
$\hl(t)$ ensures that the thermal fluctuation doesn't destroy the equilibrium of the environment. 
\end{enumerate}
\section{Dynamical equation of the system}
\label{sec-III}
It is already known that if such a system is initially oriented in the `$X$' direction, on-resonance periodic excitation leads to a long-lived prethermal plateau \cite{saha2023}. To explain such dynamics, we need to go beyond the unitary quantum evolution.  A clear understanding of the relevant timescales and the relaxation processes is also required in this case. There exists a timescale separation in this system, i.e., bath correlation time is much shorter than the system relaxation time, $\tau_c < T_1, T_2$. Here, $T_1, T_2$ are the system relaxation time scales. The system also evolves through the following relaxation processes: drive-induced dissipation (DID) \cite{chakrabarti2018b}, dipolar relaxation \cite{saha_effects_2022, saha2022}, and system-bath relaxation \cite{breuer2002}. A typical Born-Markov master equation could not explain the DID. On the other hand, the Redfield equation does not apply to the solid-state NMR \cite{redfield1957}. It is noted that a recently formulated fluctuation-regulated quantum master equation (FRQME) performs better in this case to describe the underlying physics of the spin-locking process \cite{chakrabarti2018b}. Such a formalism also gets lots of success in explaining other phenomena in driven dissipative quantum systems \cite{chanda2020,chanda2021, chatterjee_nonlinearity_2020, gourab2024, Chatterjee2024, chanda2023,fency2025, Saha_2025}.  Hence, we use FRQME to derive the dynamical equation of the system. 

\subsection{Fluctuation-regulated quantum master equation (FRQME)}
The analytical form of FRQME in the interaction 
picture of the free Hmiltonian $\hs^{\circ} + \hl^{\circ}$ is given by \cite{chakrabarti2018b},
\begin{eqnarray} \label{frqme}
\frac{d\rh}{dt}&=& -i \trl\Big[\heff(t),\rh \otimes\rl\Big]^{\rm sec}\nn\\
&&-\int\limits^{\infty}_0 d\tau \trl\Big[\heff(t),\Big[\heff(t-\tau),\rh \otimes\rl\Big]\Big]^{\rm
sec}e^{-\frac{\vert\tau\vert}{\tau_c}},
\label{eq:2}
\end{eqnarray}
$\heff(t)$ is the interaction representation of $\hsl +\hdr(t)+\hdd  + \hcs$.  `sec' denotes the secular
approximation. The presence of the exponential kernel $\left(\exp(-t/\tau_c)\right)$ in \emph{r.h.s} gives a finite second-order effect of any local Hamiltonian (e.g.
$\hdr(t)$, $\hdd$, and $\hcs$) along with $\hsl$, which plays a pivotal role in our case. 
In the interaction frame, $\hdr(t)$ and $\hdd$ can be written as a sum of secular and non-secular parts. The secular (non-secular) part is defined as the time-independent (dependent) part of $\heff(t)$ in that particular frame. As such, $H_{\scriptscriptstyle \rm DR}^{\rm sec} = \omega_1 \sum\limits_i I_x^i$ and $H_{{\scriptscriptstyle \rm DD}}^{\rm sec} =  \omega_{d_0}\mathcal{T}^{0}_2$. Similarly,  $H_{{\scriptscriptstyle \rm DR}}^{\rm n}  = \sum\limits_{i}\omega_1 \left(
I_+^i e^{+2i\omega_{\circ}t} + I_-^i e^{-2i\omega_{\circ}t} \right)$, $H_{{\scriptscriptstyle \rm DD}}^{\rm n} =\sum\limits_{m }  (-1)^m \omega_{d_{m}}
\mathcal{T}^{m}_2e^{-i m \omega_{\circ}t}\quad[\forall m\neq0]$. 

We will follow a Liouville space prescription to solve the dynamical equation numerically.
In Liouville space, the dynamical equation can be written as $\frac{d \hat{\rh}(t)}{dt}=
\hat{\mathcal{L}}\hat{\rh} (t)$. We note that, $\hat{\mathcal{L}}$ consists of three terms,
$\hat{\mathcal{L}} = \left(\hat{\mathcal{L}}_{sec}  + \hat{\mathcal{L}}_{n}  +
\hat{\mathcal{L}}_{\text{\tiny{SL}}}\right)$. They correspond to the Liouvillian corresponding to the
secular parts, the non-secular parts, and the system-bath coupling Hamiltonian, respectively.   The analytical form of the individual terms is given by
\begin{enumerate}
    \item $\mathcal{L}_{sec}\rh(t)= -i\left[\mathcal{H}_{tot}^{sec}, \rh(t)\right] -  \tau_c\left[ \mathcal{H}_{tot}^{sec},\left[\mathcal{H}_{tot}^{sec},\rh(t) \right] \right]$.
    
    Here, ${H}_{tot}^{sec} =  H_{\scriptscriptstyle \rm DR}^{\rm sec} +  H_{\scriptscriptstyle \rm DD}^{\rm sec} + \hcs$
    \item  
$\mathcal{L}_{n}\rh=  \sum\limits_{m=-2}^2 \vert\omega_{d_m}\vert^2 \tau_c\mathcal{Z}(m) \left[2 \mathcal{T}_2^{-m} \rh \mathcal{T}_2^{m} - \left\{\mathcal{T}_2^{m}\mathcal{T}_2^{-m}, \rh \right\}\right]\quad [\forall m \neq 0]$ 
Here, $\mathcal{Z}(m) = 1/(1 + (m \omega_\circ \tau_c)^2)$
\item  
$\mathcal{L}_{\text{\tiny{SL}}}\rh = \sum\limits_{i=1}^2P_{\mp}\left[  2I_{\mp}^i \rh I_{\pm}^i - \{I_{\pm}^i I_{\mp}^i, \rh\}
\right]$, 

Here $P_\pm \pm i\delta \omega_\pm = \int_0^\infty d\tau \wsl^2 e^{-\tau/\tau_c} e^{\mp i  (\omega_L - \omega_{\circ})\tau}\trl \{L_\pm L_\mp \rl\}.$
 We also assume that $\trl \{ L_\pm L_\mp \rl \} = 1 \pm \mth$. Hence, the system will eventually
inherit the equilibrium populations $(1   \pm \mth)/2$ in the absence of other local interactions at the steady state. 
 $P_-, P_+$ are defined as the downward and upward transition probabilities.
We define,  $P_+ + P_- = 1/T_1$.  
\end{enumerate}
In this prescription, we neglect all the other shift terms (e.g., Lamb shift, Bloch-Siegert shift, etc) as they have no significant contribution to the dynamics.
The two spin density matrices can also be written as a sum of relevant observables, which are given by,
\begin{eqnarray}\label{obs}
\rh (t) &=& \sum\limits_{\alpha^\prime, \beta ^\prime}  A_{\alpha^\prime \beta^\prime} I_{\alpha^\prime} \otimes I_{\beta^\prime},
\end{eqnarray}
Here, $\alpha^\prime$, $\beta^\prime$ can take values from $\{x,y,z,d\}$, and $I_d = 2\times 2$ identity matrix.
In terms of observables, the Lindblad equation can be represented as linear first-order coupled differential equations. A detailed form of the differential equations can be found in our previous works. In the next section, we show the solution of the relevant observables to explore the use of spin-locking as a chemical identifier.
\subsection{ Different relaxation times of the system}
Depending on the strength of the three terms in the Liouvillian, different time scales exist in the system. Following the experimental parameters, we are working in the regime, $\vert  \mathcal{L}_{Sec} \vert >\vert  \mathcal{L}_{n} \vert >\vert  \mathcal{L}_{\text{\tiny{SL}}}\vert$. We note that such a regime naturally arises in solid-state NMR. The corresponding relaxation times are given by,
 \begin{enumerate}
     \item The terms corresponding to $\vert  \mathcal{L}_{sec} \vert$ decay at a much faster time scale, the expression for the relaxation rate is given by,  $\omega_{\rm sec}^2 \tau_c$. Here, $\omega_{\rm sec}$ is the average strength of the local interaction. 
     \item The terms corresponding to $\vert  \mathcal{L}_{n} \vert$ decays at relatively longer  time scale. The relaxation rate is given by,  $\omega_{\rm sec}^2 \tau_c/(1 + \omega_\circ^2 \tau_c^2)$.
     \item Finally $\vert  \mathcal{L}_{\text{\tiny{SL}}}\vert$ decays with the longest timescale and the corresponding rate is given by, $\wsl^2 \tau_c/(1 + \omega_\circ^2 \tau_c^2)$.
 \end{enumerate}
\subsection{ Numerical solution of the dynamical equation}
\begin{figure}[htbp]
    \centering
    \includegraphics[width=0.5\linewidth]{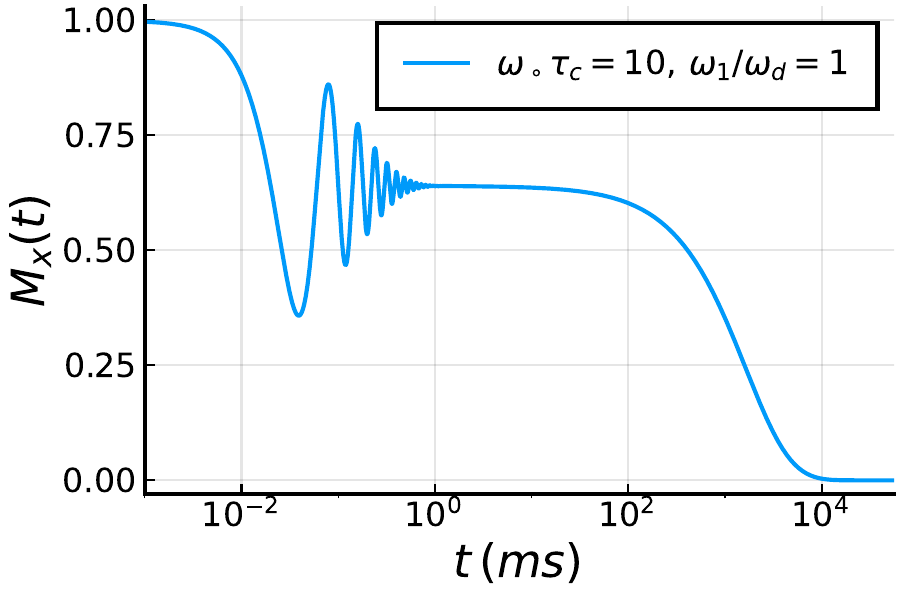}
    \caption{Depicts the plot of $M_x(t)$ vs t. The values of parameters is given by, $\omega_1, \omega_d = 2\pi \times 5$ kHz, $\tau_c = 10^{-3}$ ms. For an initial choice of $M_x(0) =1$, the dynamics show the notion of prethermalization. As such, the system reaches a prethermal quasi-steady state $M_x(t_{\rm pre})$, before it reaches the thermal state on a longer timescale.}
    \label{fig:mx1}
\end{figure}

Details on the dynamical equation in terms of observables can be found in the appendix \ref{app}. The analytical solution of $M_x(t)$ is written as,
\begin{eqnarray}
    M_x(t) = M_x(0) \left(\frac{4 \omega_1^2}{\kappa^2} + \frac{9 \omega_d^2}{4\kappa^2} \cos(\kappa t)e^{-\kappa^2 t \tau_c }  \right) e^{- \kappa_\prime^{2 }\tau_c  t}
    \label{mx(t)}
\end{eqnarray}
Here $\kappa^2  = 4 \omega_1^2 + 9\omega_d^2 /4$. and $\kappa_\prime = (\frac{5}{2}\vert \omega_{d_1}^2 \vert \mathcal{Z}(1) + \vert\omega_{d_2}^2 \vert \mathcal{Z}(2) )$. To provide a simplified picture of the prethermal state, we put $\alpha=0$. 
The key features of the equation are given by  
\begin{enumerate}
    \item At a timescale $T_{\rm pre} = 1/\kappa^2\tau_c$, the system reaches a quasi-steady state (i.e., prethermal state). 
    \item The corresponding value of $M_x(t_{\rm pre}) = M_x(0) \frac{16 \omega_1^2}{16 \omega_1^2 + 9 \omega_d^2}$. We note that the prethermal value is independent of the environmental parameter.
    \item At a time  $T_{\ rmth}=1/\kappa_{\prime}{^2 }\tau_c$, the prethermal state decays and the system reaches the constrained thermal state.
\end{enumerate}
We numerically solve the dynamical equation and plot the transverse magnetization along the `$X$' axis (i.e., $M_x$). The initial magnetization is chosen along the same axis, as such $M_x(0) = 1$. Following the experimental choice of  parameters \cite{beatrez_floquet_2021}, $\omega_1 \geqslant \omega_d>\wsl$, and $\omega_\circ \tau_c>1$, we show the dynamics of $M_x(t)$ vs $t$ in Fig. \ref{fig:mx1}, where the emergence of prethermal state is clearly visible. 
For the above case, the effect of $\hsl$ and the counter-rotating terms of on-resonant drives are also neglected for simplicity. 

The prethermal state can be characterized by two quantities, 
\begin{enumerate} 
    \item The value of the prethermal state $M_x(t_{\rm pre})$.
    \item The lifetime of the prethermal state (i.e., $T_{\rm th} - T_{\rm pre} $).
\end{enumerate}  
We note that both of the quantities can be measured by typical experiments in solid-state NMR \cite{beatrez_floquet_2021}. In the next section, we will use these two parameters to use such pulse techniques as an identifier of the chemical components. 

\section{Application in the characterization of local environments}
\label{sec-IV}
In this section, we show that using the spin-locking pulse sequence, the presence of several chemical components can be detected. Among them, we only focus on two types of chemical components, which are given below.

\subsection{Characterizing the dynamics of the environment }
It is helpful in several contexts to detect the chemical ingredients of the external environment. Previous schemes are related to $T_1,\, T_2$ measurement. $T_1,\, T_2$ depend on the bath spectral density function. The spectral density directly depends on $\tau_c$. Hence, measuring the relaxation time was efficient in predicting the fluctuation correlation time \cite{Duer2002}.

 Here we show that using spin-locking, we can also detect the environmental parameter $\tau_c$.  As we note that $M_x(t_{\rm pre})$ is $\tau_c$ independent, hence the lifetime of the prethermal state would be an identifier for that.
\begin{figure}[htbp]
    \centering
     \raisebox{3cm}{\normalsize{\textbf{(a)}}}\hspace*{-1mm}
    \includegraphics[width=0.4\linewidth]{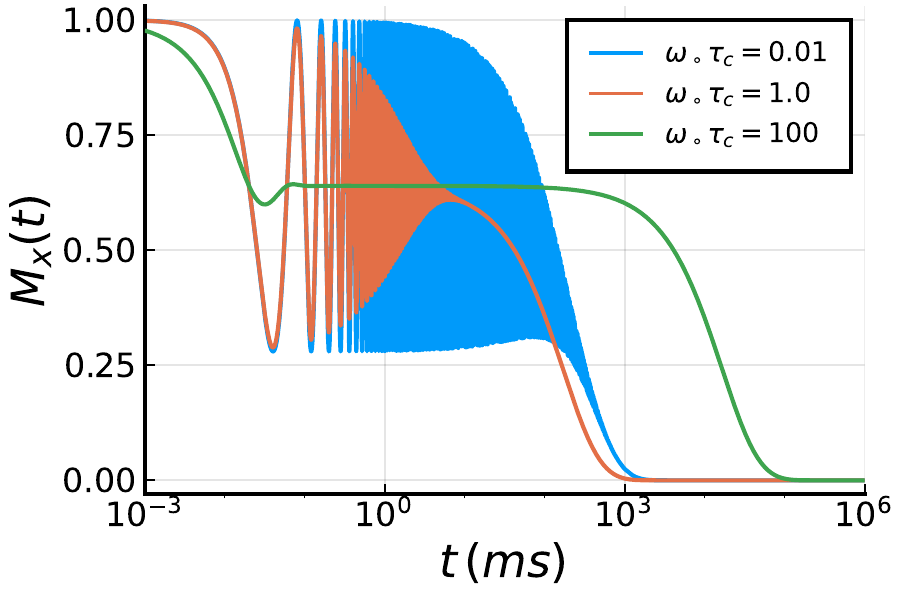}
     \raisebox{3cm}{\normalsize{\textbf{(b)}}}\hspace*{-1mm}
    \includegraphics[width=0.4\linewidth]{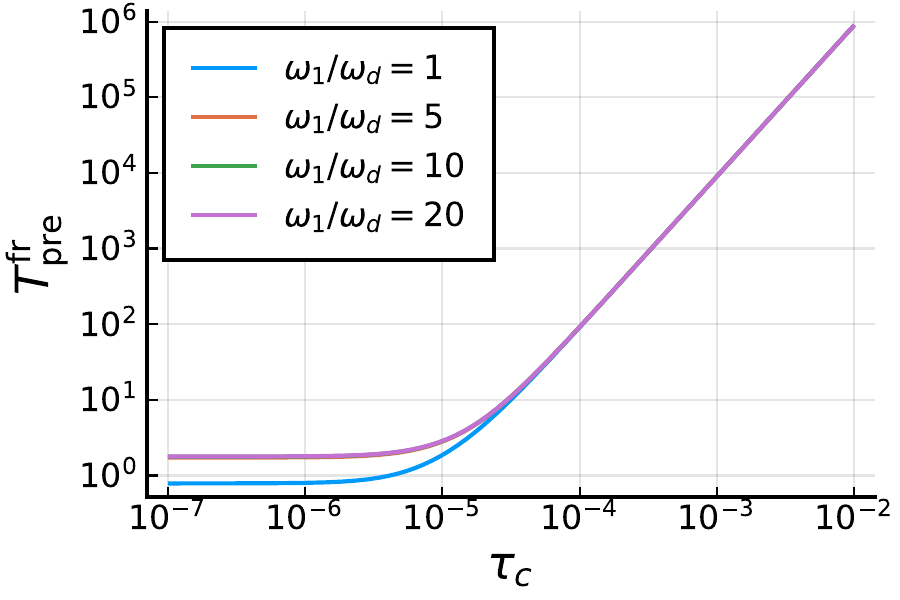}
    \caption{(a) shows the plot of $M_x(t)$  vs $t$ for several values of $\tau_c$. Here $\omega_1, \omega_d = 2\pi \times 5$ kHz, $\omega_\circ = 2\pi \times 10^4$ kHz. For decreasing the value of $\tau_c$, the lifetime of the prethermal state diminishes. (b) show the plot the fractional lifetime of prethermal state, $T_{\rm pre}^{\rm fr}$ (which is defined in Eq. (\ref{lifetime}) vs $\tau_c$ for different values of $\omega_1/\omega_d$. For $\omega_\circ \tau_c <1$ (amorphous solid), $T_{\rm pre}^{\rm fr}$  has no dependence in $\tau_c$. However, $\omega_\circ \tau_c>1$ (crystalline solid) has a linear dependency. Hence, the lifetime of the prethermal state is a good identifier of the chemical properties of the environment. }
    \label{fig:mx2}
\end{figure}
we show that for changing $\omega_\circ \tau_c$, the lifetime of the prethermal state modifies. As such, the prethermal state never appears for $\omega_\circ \tau_c <1$. Hence, $\tau_c \approx 1/\omega_\circ$ is also known as the critical point \cite{saha2023}. We plot the lifetime of the prethermal state by varying $\tau_c$ in Fig. \ref{fig:mx2}(a). We define the fractional prethermal time as, 
\begin{eqnarray}
 T_{\rm pre}^{\rm fr} = \frac{T_{\rm th} - T_{\rm pre}}{T_{\rm pre}}  
 \label{lifetime}
\end{eqnarray}
We show that $T_{\rm pre}^{\rm fr}$ sharply decays for decaying $\tau_c$ in Fig. \ref{fig:mx2}(b). Hence, it could be a good identifier for $\tau_c$. The key result of this section follows,
\begin{enumerate}
    \item If such a system doesn't show the notion of prethermalization, then $\tau_c < 1/ \ 1/\omega_\circ$. Such a condition is only valid for a chemical environment that consists of amorphous solids.
    \item On the other limit, $\tau_c > 1/ \omega_\circ$, The value of $T_{\rm pre}$ can be used for describing the properties of the chemical environment. Such a condition is valid for crystalline solids.
    \item The ratio of $\omega_1/\omega_d$ doesn't matter for the prethermal state lifetime in the regime of $\tau_c > 1/\omega_\circ$. Hence, such a ratio does not affect the prediction of $\tau_c$.
\end{enumerate}

\subsection{Characterizing the dynamical shifts }
 In the previous case, the lifetime of the prethermal state is used for our analysis as the prethermal magnetization has no dependence on $\tau_c$. However, $\alpha$ affects both lifetime and the value of $M_x(t_{\rm pre})$.  In this section, we use the latter one for the characterization of chemical shifts.
\begin{figure}[htbp]
    \centering
    \raisebox{3cm}{\normalsize{\textbf{(a)}}}\hspace*{-1mm}
    \includegraphics[width=0.5\linewidth]{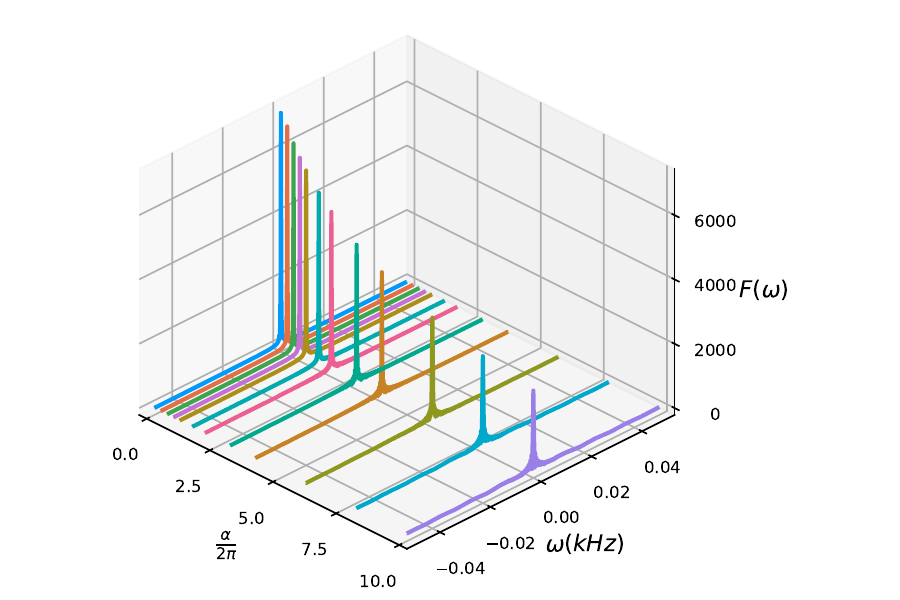}
    \raisebox{3cm}{\normalsize{\textbf{(b)}}}\hspace*{-1mm}
    \includegraphics[width=0.4\linewidth]{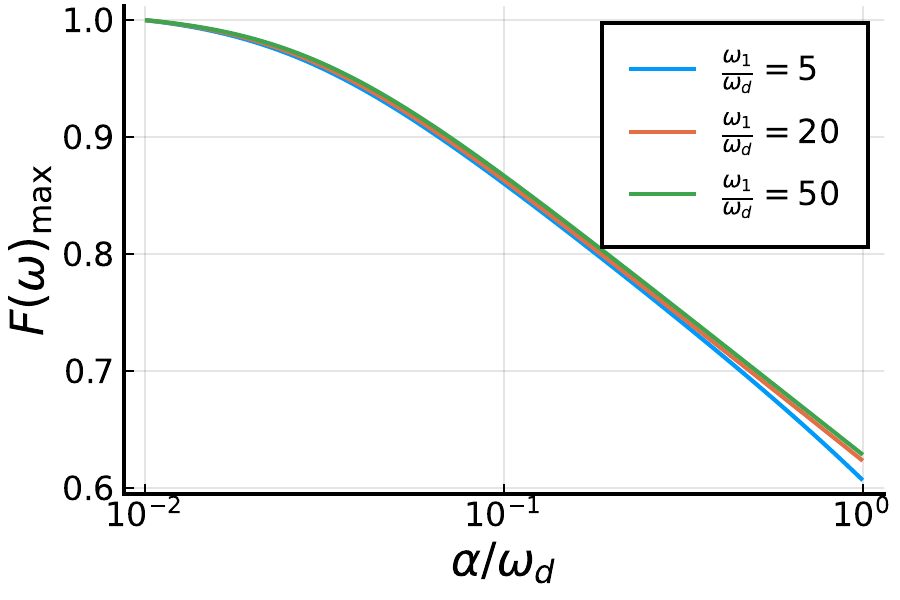}
    \caption{(a) shows the 3D plot of Fourier transform of $M_x$, $\mathcal{F}(\omega)$ vs chemical shift strength $\alpha$ vs $\omega$. The Fourier peak arises at $\omega=0$, as we consider the on-resonant periodic drive. Here $\omega_1 = 2\pi \times 5$ kHz, $\omega_d = 2\pi \times 5$ kHz, $\omega_{\circ} = 2\pi \times 10^4$ kHz, and $\tau_c = 10^{-4}$ ms. The Fourier peak gradually decreases with increasing shift strength. In (b), we show the plot of the height of the Fourier peak of $\mathcal{F}(\omega)$ vs $\alpha$ for different values of $\omega_1/\omega_d$. We have two observations: the ratio of $\omega_1/\omega_d$ has no such significant dependence on the plot, and for increasing $\alpha$, the peak height decreases; hence, the value of $M_x(t_{\rm pre})$ is useful for the characterization of the chemical shift.}
    \label{fig:mx3}
\end{figure}
 The presence of chemical shift plays a destructive role in the prethermal state. We note that $\hcs$ doesn't commute with the observables corresponding to the conserved quantity for the prethermalization (shown in Appendix- \ref{app}), hence $M_x(t_{\rm pre})$ will further decay due to $\alpha$. The change in $M_x(t_{\rm pre})$ could be useful for depicting the chemical properties of $\hcs$. We plot the Fourier transform $F(\omega)$ of $M_x(t)$ \emph{vs} $\omega$, for different values of $\alpha$ in Fig \ref{fig:mx3}(a), (b). Our numerical result matches the theoretical understanding, where the peak of the Fourier spectrum decreases for increasing $\alpha$. It is also noted that for larger values of $\alpha$ (i.e.,$\alpha>\omega_1,\omega_d$), no such prethermalization occurs; hence, we confine ourselves to the regime $\alpha \leqslant \omega_d$. A detailed discussion on the dependence of lifetime on the chemical shift is explained in the next session.
\section{Discussion}
\label{sec-V}
For a good characterization of the chemical components, the process or pulse technique must be able to detect small changes in those elements. We show that the spin-locking fulfills the criteria. Previous works are concentrated on the measurement of $T_{1 \rho}$ \cite{Almeida2024}. Here, we choose a different path. Recently, prethermalization using spin-locking has been explored in several contexts \cite{beatrez_floquet_2021, Beatrez2023,saha2023, Saha2024}. Hence, we decide to provide a characterization protocol using the prethermal state. We find that the value of the traverse magnetization at the prethermal state and the prethermal state lifetime is a good identifier for that. 

We focus on the two cases, (i) the chemical components of the external environment and
(ii) the presence of any chemical or dynamical shifts. For characterizing the local
environments, we find that the lifetime of the prethermal state is efficient. We define
the lifetime of the prethermal state in Eq. (\ref{lifetime}) and plot it as a function of
$\tau_c$. We show that the environment has a very short fluctuation timescale $(\tau_c <
1/\omega_\circ)$ (e.g., amorphous solids), then there is no existence of a prethermal
state. On the other limit $(\tau_c > 1/\omega_\circ)$ (e.g., crystalline solids), the
lifetime of the prethermal state grows with increasing fluctuation-time scale (shown in
Fig. \ref{fig:mx2} ). As such, for a crystalline environment, the lifetime could be
improved by freezing or reducing the degrees of freedom. Hence, from the value of
$\tau_c$, the properties of the local environments can easily be identified.

Similarly, our theoretical results will also be helpful in depicting chemical shifts. We note that for a stronger shift  (i.e., stronger than the dipolar interaction), our methods will not work as the prethermal state will thermalize quickly. For a weaker shift, we show that the value of $M_x$ changes when the shift strength is changed. Therefore, measuring the value of $M_x$ or $\mathcal{F}(M_x)$ will be helpful in the characterization of the chemical shift.
\begin{figure}[htbp]
    \centering
    \includegraphics[width=0.5\linewidth]{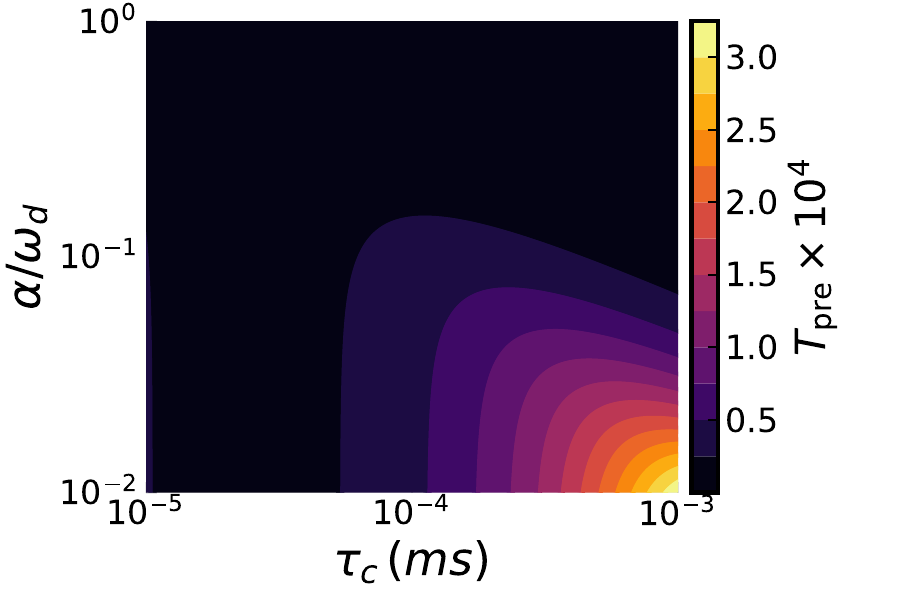}
    \caption{depicts the contour plot of $T_{\rm pre}$ (defined in Eq. (\ref{ana-life})) as a function of the relative strength of chemical shift $\alpha/\omega_d$ and environmental correlation time $\tau_c$. the lifetime increases for increasing $\tau_c$ and decreasing $\alpha/\omega_d$.  The contour plot indicates that the prethermal state can be used for the characterization of chemical shifts and the environment. }
    \label{contour}
\end{figure}
We also show the contour plot of the lifetime of the prethermal state by varying $\alpha$ and $\tau_c$.  An alternative theoretical definition of such a lifetime is provided, which can be used for the numerical verification of those experimental outcomes. The dynamical equation has the form, $\dot{ \rho} = \mathcal{L} \rho$. We note that $ \mathcal{L}$ always has an eigenvector with a zero eigenvalue, which ensures the trace preservation and the existence of a steady state (i.e., thermal state). In the case of prethermalization, there exist other eigenvectors whose eigenvalues are close to zero compared to the remaining ones \cite{saha2024prethermalization}. The real part of the corresponding eigenvalues represents the decay rate. Hence, the lifetime of the prethermal state can be defined as,
\begin{eqnarray}
    T_{\rm pre} = 1/\vert\mathcal{L}_{NNN} - \mathcal{L}_{NN}\vert.
    \label{ana-life}
\end{eqnarray}
Here, $NN$ and $NNN$ represent the nearest neighbor and next nearest neighbor, respectively. $1/\vert\mathcal{L}_{NNN}\vert$ is the timescale for the creation of the prethermal state, and $1/\vert\mathcal{L}_{NN}\vert$ gives the notion of decay time of the prethermal state.

Using this definition, we plot $T_{\rm pre}$ as a function of $\alpha$ and $\tau_c$ in
Fig. \ref{contour}. We show that higher values of $\tau_c$ and lower values of $\alpha$
provide a long-lived prethermal state. Hence, the lifetime of the prethermal state will be
a good quantity for characterizing the environment and the shift. We note that the recent
developments in solid-state NMR to explore the prethermal state and its properties can
also be helpful to verify our theoretical prediction \cite{beatrez_floquet_2021,
Beatrez2023, Beatrez2023b}. 

In the manuscript, we only study a homonuclear system consisting of two dipolar-coupled
spins to exhibit several uses of a prethermal state.  An extension of our analysis can
also be done for the heteronuclear case. We hope that the extra detuning term will be
effective in identifying the chemical components. On the other hand, a rigorous study
considering the anisotropic part of the chemical shift is required for the detailed
analysis of the prethermal state in the presence of chemical shift. Similarly, we like to
verify our claims for a real experimental setup of a dissipative quantum many-body system,
where the dipolar-coupled spins are randomly oriented. The orientation of the spins and
the geometry of the system may play a vital role. Another open question would be to
predict the lifetime for a non-Markovian bath, where the chemical environment has some
finite memory, i.e., instead of a delta correlation, which we assume in this case, there
would be a Gaussian or Lorentzian correlation.

\section{Conclusion}
\label{sec-VI}
We revisit the theory of prethermalization using a spin-locking pulse in solid-state NMR.
We have used the recently formulated FRQME for the analysis. We predicted that such a
prethermal state would be helpful in characterizing the local environment. The prethermal
state can be specified by its lifetime and the value of transverse magnetization. We show
that the lifetime could be a good identifier for the characterization of the chemical
environment. Similarly, transverse magnetization can also be used to identify the presence
of weak chemical shifts. We envisage that this method will be a valuable addition to the existing arsenal of
solid state NMR methods.

\section{Acknowledgments}
We dedicate this article to the memory of Prof. Dr. Anil Kumar, an eminent scientist and a
pioneer in the field of Nuclear Magnetic Resonance (NMR) in India. His groundbreaking
contributions, mentorship, and vision have left a lasting impact on the scientific
community. This work stands as a humble tribute to his legacy. S.S. sincerely thanks
Sarfraj Fency, Sheelbhadra Chatterjee, and Ipsita Chakraborty for helpful discussions.

\ Appendix
    \label{app}
    \section{Dynamical equations in terms of observables}
Here we additionally show the dynamical equations in terms of observables defined in Eq. (\ref{obs}). 
\subsection{Dynamical equations for $\hat{\mathcal{L}}_{sec}$  }
Here we show only the equations that are connected to $M_x(t)$. Others are neglected as they have negligible contributions to $M_x$.
\begin{eqnarray}
\dot{M}_{x} &=& -\frac{9}{4} \omega_{d_0}^2 \tau_c  M_{x} + 6 \omega_1 \omega_{d_0} \tau_c M_{zz} -6 \omega_1 \omega_{d_0} \tau_c M_{yy} -3\omega_{d_0} M_{yz} \label{eq-pre1} \\
\dot{M}_{zz} &=&  \frac{3}{4}\omega_1 \omega_{d_0} \tau_c M_{x} -2\omega_1^2 \tau_c M_{zz} +  2\omega_1^2 \tau_c M_{yy} + \omega_1 M_{yz}\label{eq-pre2} \\
\dot{M}_{yy} &=&  -\frac{3}{4}\omega_1 \omega_{d_0} \tau_c M_{x} +2\omega_1^2 \tau_c M_{zz} -  2\omega_1^2 \tau_c M_{yy} - \omega_1 M_{yz}\label{eq-pre3} \\
\dot{M}_{yz} &=& \frac{3}{4}\omega_{d_0} M_{x} -2 \omega_1 M_{zz} +2\omega_1 M_{yy} -(4 \omega_1^2 + \frac{9}{4}\omega_{d_0}^2 \tau_c)M_{yz} \label{eq-pre4}
\end{eqnarray}

Several conserved quantities exist in the dynamics.  The analytical forms are given as,
$M_{yy}+M_{zz}$, $3\omega_{d_0}M_{zz}+\omega_1  M_{x}$, and $M_{xx}$. The Liouvillian has four zero eigenvalues (i.e., trace preservation and three conserved quantities mentioned above). The chemical shift is assumed to be zero for this case. We note that for adding the shifts, $3\omega_{d_0}M_{zz}+\omega_1  M_{x}$, $M_{yy}+M_{zz}$, and $M_{xx}$ will not be conserved anymore. However, $M_{xx}+M_{yy}+M_{zz}$ will be conserved throughout the dynamics (we exclude the effect of $\hsl$ here).

\subsection{Dynamical equations for $\hat{\mathcal{L}}_{n}$ }
The dynamical equation in terms of observables is given as,
\begin{eqnarray}\label{eq-4}
\dot{M}_x &=& - (  \frac{5}{2}p_1 + p_2) M_x \label{eq-ct1}\\
\dot{M}_{zz} &=& -   2 p_1 M_{zz} + p_1(M_{xx} + M_{yy})  \label{eq-ct2}\\
\dot{M}_{yy}&=& -( p_2 + p_1)M_{yy} + p_2 M_{xx} +    p_1 M_{zz}  \label{eq-ct3}\\
\dot{M}_{xx}&=& - ( p_2 + p_1)M_{xx} + p_2 M_{yy} +    p_1 M_{zz} \label{eq-ct4}
\end{eqnarray}
Here  $p_m = \vert\omega_{d_m}\vert^2 \tau_c\mathcal{Z}(m) $. 
There is a conserved quantity, which is given by $M_{xx} + M_{yy} + M_{zz}$. Similarly 
$\hat{\mathcal{L}}_{n}$ has two zero eigenvalues (i.e., trace preservation and the above-mentioned conserved quantity). The conservation law will survive under the influence of $\hcs$.

\bibliographystyle{apsrev4-1}
\bibliography{ref2}
\end{document}